\def\be{\begin{equation}}
\def\ee{\end{equation}}
\def\ba{\begin{array}}
\def\ea{\end{array}}

\documentclass[aps,amsmath,amssymb,amsfonts]{revtex4}
\usepackage{mathrsfs}
\usepackage{graphicx}
\usepackage{epstopdf}
\def\qed{\leavevmode\unskip\penalty9999 \hbox{}\nobreak\hfill
     \quad\hbox{\leavevmode  \hbox to.77778em{%
               \hfil\vrule   \vbox to.675em%
               {\hrule width.6em\vfil\hrule}\vrule\hfil}}
     \par\vskip3pt}

\begin{document}
\title{General Monogamy and polygamy properties of quantum systems}

\author{Bing Xie}
\affiliation{Department of Mathematics, East China University of Technology, Nanchang 330013, China}

\author{Ming-Jing Zhao}
\affiliation{
School of Science, Beijing Information Science and Technology University, Beijing, 100192, PR China}
\author{Bo Li\footnote{Corresponding author}}
\email{libobeijing2008@163.com.}
\affiliation{School of Computer and Computing Science, Zhejiang University City College, Hangzhou 310015, China}

\begin{abstract}
Monogamy and Polygamy are important properties of entanglement, which characterize the entanglement distribution of multipartite systems. We study general monogamy and polygamy relations based on the $\alpha$th $(0\leq\alpha\leq \gamma)$ power of entanglement measures and the $\beta$th $(\beta\geq \delta)$ power of assisted entanglement measures, respectively. We illustrate that these monogamy and polygamy relations are tighter than the inequalities in the article [Quantum Inf Process
19, 101], so that the entanglement distribution can be more precisely described for entanglement states that satisfy stronger constraints. For specific entanglement measures such as concurrence and the
convex-roof extended negativity, by applying these relations, one can yield the corresponding monogamous and polygamous inequalities,
which take the existing ones in the
articles [Quantum Inf Process 18, 23] and [Quantum Inf Process 18, 105] as special cases. More details are presented in the examples.
\end{abstract}

\maketitle

\section{Introduction}
As a physical resource, quantum entanglement \cite{s1,s2,s3,s4,s5,s6} is of great research significance and is widely used in areas such as quantum teleportation \cite{s7}, quantum key distribution \cite{s8}, and quantum computing \cite{s9}. Since the concept of quantum entanglement was proposed in 1935, many endeavors have been devoted to exploring how to quantify quantum entanglement. The earliest quantitative studies were on bipartite qubit systems, where the main entanglement measures were concurrence \cite{s10}, entanglement of formation \cite{s11} and negativity \cite{s12}. With the deepening of research, it is found that the distribution of quantum information resources is unequal \cite{s13}. This phenomenon is known as entanglement monogamy. Specifically, this phenomenon shows that for a multipartite qubit system, the entanglement of one subsystem with another will limit the entanglement of this subsystem with the rest of the subsystems.

Monogamy describes the distribution of entanglement in multipartite qubit systems, furthermore, it is one of the important ways to study the structure and properties of entanglement. For a tripartite state $\rho_{A|BC}$, Coffman, Kundu, and Wootters first proposed that the entanglement $\mathcal{E}$ between $A$ and $BC$ follows a monogamy relation $\mathcal{E}_{A|BC} \geq \mathcal{E}_{AB}+ \mathcal{E}_{AC}$, commonly known as the CKW inequality \cite{s13}. However, entanglement measures always violate such monogamy relation. In fact, it has been proved that the squared concurrence $C^2$ \cite{s14} and the squared entanglement of formation $E^2$ \cite{s15,s16} satisfy the monogamy relations for multipartite qubit systems. The authors in Ref. \cite{s17} proved that the $\alpha$th power of concurrence $C^\alpha$ for $\alpha\geq 2$ and the $\alpha$th power of entanglement of formation $E^\alpha$ for $\alpha\geq \sqrt{2}$ satisfied the general monogamy property. According to the CKW inequality, the monogamy relation was generalized to other entanglement measure such as convex-roof extended negativity $\widetilde{N}$ \cite{s18,s19,s20}. Moreover, it has been shown that these entanglement measures to the $\alpha$th power do satisfies tighter monogamy relations of tripartite systems and multipartite systems \cite{s23,s24,s25,s26}.

The assisted entanglement is the dual concept of entanglement for bipartite systems. For any tripartite pure state $|\psi_{ABC}\rangle$, the assisted entanglement $\mathcal{E}_a$ is defined as \cite{s28}
$$\mathcal{E}_a(|\psi_{ABC}\rangle)  \equiv \mathcal{E}_a(|\rho_{ABC}\rangle)  \equiv \max_{\{p_i,|\phi_i \rangle\}} \sum_{i}p_i\mathcal{E}(|\phi_i\rangle), $$
where the maximum is taken over all possible decompositions of $\rho_{AB}=\sum_{i}p_i|\phi_i\rangle$.
As another entanglement constraint in multipartite entanglement, polygamy can be described by assisted entanglement. For arbitrary tripartite systems, Gour \emph{et.al} in Ref. \cite{s27} established the first polygamy inequality by using the squared concurrence of assistance $C_a^2$. For a tripartite state $\rho_{A|BC}$, polygamy of entanglement is characterized by $\mathcal{E}_{aA|BC} \leq \mathcal{E}_{aAB}+ \mathcal{E}_{aAC}$ with the assisted entanglement $\mathcal{E}_a$ between $A$ and $BC$. This polygamy relation has been generalized to multipartite systems for various assisted entanglement measures such as the squared concurrence of assistance $C^2_a$ \cite{s28}, entanglement of assistance $E_a$ \cite{s17,s29}, the $\beta$th $(0\leq \beta \leq 2)$ power of convex-roof extended negativity of assistance $\widetilde{N}^\beta_a$ \cite{s30}. Similar to the monogamy relation, the corresponding tighter polygamy relations have also been established for
different kinds of assisted entanglements \cite{s22,s25,s26}.

Recently, for the existing tighter inequalities in Refs. \cite{s22,s24,s25}, the corresponding monogamy relations of the $\alpha$th power of entanglement measure and the corresponding polygamy relations of the $\beta$th power of assisted entanglement measure have been proved for different parameter ranges \cite{s33,s34,s35}. However, the relations in Ref. \cite{s26} with different parameter ranges is unknown. In this paper, based on the previous results, we obtain general monogamy inequalities of any entanglement measure satisfying $\mathcal{E}_{A|BC}^\gamma \geq \mathcal{E}_{AB}^\gamma +  \mathcal{E}_{AC}^\gamma$, so that the entanglement distribution can be described more accurately for quantum states that satisfy the stronger constraints. Additionally, we also present general polygamy inequalities in terms of any assisted entanglement measure satisfying $\mathcal{E}_{aA|BC}^\delta \leq \mathcal{E}_{aAB}^\delta +  \mathcal{E}_{aAC}^\delta$. For the same measure, the corresponding monogamy and polygamy inequalities complement the existing ones \cite{s26}, which have different regions for the parameters $\alpha$ and $\beta$, respectively. If these general inequalities are applied to specific measures, we show that the resulting inequalities are tighter than those already in Refs. \cite{s33,s34,s35}, and that our results reduce the existing inequalities when the parameters take certain values. We will illustrate these advantages with concrete examples, such as the corresponding monogamy relations for concurrence and polygamy relations for the convex-roof extended negativity.

\section{Preliminary knowledge}
In this section, we provide some preliminary knowledge, we mainly introduce the definitions of relevant entanglement measures.
Let $H_A$ and $H_B$ denote a finite dimensional complex inner product vector space associated with quantum subsystem $A$ and $B$, respectively. For a bipartite pure state $|\psi\rangle_{AB}\in H_A\otimes H_B$, the concurrence is defined as follows \cite{s36}
\begin{eqnarray*}\label{l1}
	C(|\psi\rangle_{AB})=\sqrt{{2\left[1-\mathrm{Tr}(\rho_A^2)\right]}},
\end{eqnarray*}
where $\rho_A$ is the reduced density matrix by tracing over the subsystem $B$, $\rho_A=\mathrm{Tr}_B(|\psi\rangle_{AB}\langle\psi|)$.

The concurrence for a bipartite mixed state $\rho_{AB}$ is given by the convex roof extension
\begin{eqnarray*}\label{l0}
	C(\rho_{AB})=\min_{\{p_i,|\psi_i\rangle\}}\sum_ip_iC(|\psi_i\rangle),
\end{eqnarray*}
where the minimum is taken over all possible decompositions of $\rho_{AB}=\sum_ip_i|\psi_i\rangle\langle\psi_i|$, with $p_i\geq0$, $\sum_ip_i=1$ and $|\psi_i\rangle\in H_A\otimes H_B$. It has been proved that the concurrence of a mixed state $\rho_{AB}$ can be calculated by an analytical expression as follows \cite{s37}:
$$C(\rho_{AB})=\max\{\mu_1-\mu_2-\mu_3-\mu_4,0\},$$
where $\mu_1\geq\mu_2\geq\mu_3\geq\mu_4$ are the eigenvalues of the matrix $\sqrt{\sqrt{\rho}\widetilde{\rho}\sqrt{\rho}}$. Here, $\widetilde{\rho}=(\sigma_y\otimes\sigma_y) \rho^\ast (\sigma_y\otimes\sigma_y)$ and $\rho^\ast$ is the complex conjugation of $\rho$.

For a given state $\rho_{AB}\in H_A \otimes H_B$, the entanglement negativity is defined by
\begin{eqnarray*}
	N(\rho_{AB})=\frac{||\rho_{AB}^{T_A}||-1}{2},
\end{eqnarray*}
where $||Y||=\sqrt{YY^\dag}$ is the trace norm of matrix $Y$ \cite{s32}, $T_A(\rho_{AB})$ denotes the partial transposed matrix of $\rho_{AB}$ for subsystem $A$. For convenience, we use $||\rho_{AB}^{T_A}||-1$ to define negativity in this paper.  The negativity of arbitrary bipartite pure state $|\psi\rangle_{AB}$ is $N(|\psi\rangle_{AB})=2\sum_{i\leq j}\sqrt{\mu_i\mu_j}=(Tr\sqrt{\rho_A})^2-1$, here $\mu_i$ are the eigenvalues  of the reduced density matrix $\rho_A$ of $|\psi\rangle_{AB}$.

For arbitrary bipartite mixed states $\rho_{AB}$, the convex-roof extended negativity (CREN) is given by
$$\widetilde{N}(\rho_{AB})=\min_{\{p_i,|\psi_i \rangle\}}p_iN(|\psi\rangle_{AB}),$$
where the minimum is taken over all possible decompositions of $\rho_{AB}$. For any two-qubit mixed state $\rho_{AB}$ \cite{s25}, one has $\widetilde{N}(\rho_{AB})=C(\rho_{AB})$.

As a dual concept of CREN, the convex-roof extended negativity of assistance (CRENoA) is defined as
$$\widetilde{N}_a(\rho_{AB})=\max_{\{p_i,|\psi_i \rangle\}}p_iN(|\psi\rangle_{AB}),$$
with the maximum is taken over all possible decompositions of $\rho_{AB}$. For any two-qubit mixed state $\rho_{AB}$ \cite{s20}, one has $\widetilde{N}_a(\rho_{AB})=C_a(\rho_{AB})$.

Before giving our result, we propose the following lemma.

{[\bf Lemma 1]}.
For any real numbers $s$ and $t$, $0\leq s \leq t$, we can get
\begin{eqnarray}\label{p3}
	\left (1+s\right)^m-s^m \geq \left (1+t\right)^m-t^m,
\end{eqnarray}
for $0 \leq m\leq 1$, and
\begin{eqnarray}\label{p4}
	\left (1+s\right)^n-s^n \leq \left (1+t\right)^n-t^n,
\end{eqnarray}
for $n \geq 1$.

{\sf [Proof].}
Let $f(s,m)=\left (1+s\right)^m-s^m$ with $s\geq 0$ and $0\leq m \leq 1$. Then, $ \frac{ \partial f }{ \partial s }=m [\left(1+s\right)^{m-1}-s^{m-1}]\leq 0$. Therefore, $f(s,m)$ is a decreasing function of $s$. As $s\leq t$, we obtain $f(s,m) \geq f(t,m)=\left (1+t\right)^m-t^m$. When $s=t$, the equation holds. we can get the inequality (\ref{p4}) using a similar proof, since $f(s,n)$ is an increasing function of $s$ for $s \geq 0$ and $n \geq 1$.

For convenience, in the following we denote the entanglement of $\rho_{AB_i}$ by $\mathcal{E}_{AB_i}=\mathcal{E}(\rho_{AB_i})$ and the entanglement of $\rho_{A|B_i \cdots B_{n-1}}$ by $\mathcal{E}_{A|B_i \cdots B_{n-1}}=\mathcal{E}(\rho_{A|B_i \cdots B_{n-1}})$ for an entanglement measure $\mathcal{E}$.
We also denote the entanglement of $\rho_{AB_i}$ by $\mathcal{E}_{aAB_i}=\mathcal{E}_a(\rho_{AB_i})$ and the entanglement of $\rho_{A|B_i \cdots B_{n-1}}$ by $\mathcal{E}_{aA|B_i \cdots B_{n-1}}=\mathcal{E}_a(\rho_{A|B_i \cdots B_{n-1}})$ for an assisted entanglement measure $\mathcal{E}_a$.

\section{general Monogamy properties for entanglement measures}
Let $\mathcal{E}$ be a bipartite quantum entanglement measure. For any $2\otimes2\otimes2^{n-2}$ tripartite state $\rho_{A|BC} \in H_A\otimes H_B\otimes H_C$, assume non-negative real number $\gamma$ is the value for $\mathcal{E}^\gamma$ to satisfy inequality
\begin{eqnarray}\label{p1}
	\mathcal{E}_{A|BC}^\gamma \geq \mathcal{E}_{AB}^\gamma +  \mathcal{E}_{AC}^\gamma ,
\end{eqnarray}
that is, $\gamma \in Q=\{\eta ~|~\mathcal{E}_{A|BC}^\eta \geq \mathcal{E}_{AB}^\eta +  \mathcal{E}_{AC}^\eta$ for all  tripartite state $ \rho_{ABC} \}$.
Previous studies have shown that the monogamous relation (\ref{p1}) can be applied to bipartite entanglement measures such as concurrence \cite{s17}, entanglement of formation \cite{s21} and the convex-roof extended negativity \cite{s21}.

In the following, for the $\alpha$th power of entanglement measure $\mathcal{E}^\alpha$, we will propose a new class of monogamy relation for quantum states with some constraints according to Lemma 1, which are tighter than the inequalities in Ref. \cite{s35}.

{[\bf Theorem 1]}.
For any $2\otimes2\otimes2^{n-2}$ tripartite state $\rho_{ABC}\in H_A\otimes H_B\otimes H_C$, let real number $h$ satisfy $0\leq h\leq 1$, there exists real number $u\geq 1$,

(1) if $h\mathcal{E}_{AC}^\gamma \geq \mathcal{E}_{AB}^\gamma $, the entanglement measure $\mathcal{E}$ satisfies
\begin{eqnarray}\label{p6}
	\mathcal{E}_{A|BC}^\alpha \geq \mathcal{E}_{AB}^\alpha + \left[\left(u+h\right)^\frac{\alpha}{\gamma}-h^\frac{\alpha}{\gamma}\right] \mathcal{E}_{AC}^\alpha
\end{eqnarray}
for $0 \leq \alpha \leq \gamma$ and $\gamma \in Q$.

(2) if $h\mathcal{E}_{AB}^\gamma \geq \mathcal{E}_{AC}^\gamma$, the entanglement measure $\mathcal{E}$ satisfies
\begin{eqnarray}\label{p17}
\mathcal{E}_{A|BC}^\alpha \geq  \left[\left(u+h\right)^\frac{\alpha}{\gamma}-h^\frac{\alpha}{\gamma}\right] \mathcal{E}_{AB}^\alpha + \mathcal{E}_{AC}^\alpha
\end{eqnarray}
for $0 \leq \alpha \leq \gamma$ and $\gamma \in Q$.

{\sf [Proof].}
For any $2\otimes2\otimes2^{n-2}$ tripartite state $\rho_{A|BC}$, one has relation $\mathcal{E}_{A|BC}^\gamma \geq \mathcal{E}_{AB}^\gamma + \mathcal{E}_{AC}^\gamma$. Therefore, there exists $u \geq 1$ such that
\begin{eqnarray}\label{p5}
	\mathcal{E}_{A|BC}^\gamma \geq \mathcal{E}_{AB}^\gamma +u \mathcal{E}_{AC}^\gamma.
\end{eqnarray}
From the relation (\ref{p5}) and inequality (\ref{p3}) in Lemma 1, we have
\begin{eqnarray*}\label{l5}
	\mathcal{E}_{A|BC}^\alpha
	&&\geq (\mathcal{E}_{AB}^\gamma + u\mathcal{E}_{AC}^\gamma)^\frac{\alpha}{\gamma}\\
	&& = \mathcal{E}_{AB}^\alpha +u^\frac{\alpha}{\gamma}\mathcal{E}_{AC}^\alpha\left[\left(\frac{1}{u}\frac{\mathcal{E}_{AB}^\gamma}{\mathcal{E}_{AC}^\gamma}+1\right)^\frac{\alpha}{\gamma}-\left(\frac{1}{u}\frac{\mathcal{E}_{AB}^\gamma}{\mathcal{E}_{AC}^\gamma}\right)^\frac{\alpha}{\gamma}\right]\\
	&&\geq \mathcal{E}_{AB}^\alpha +u^\frac{\alpha}{\gamma}\mathcal{E}_{AC}^\alpha\left[\left(\frac{h}{u}+1\right)^\frac{\alpha}{\gamma}-\left(\frac{h}{u}\right)^\frac{\alpha}{\gamma}\right] \\
	&& = \mathcal{E}_{AB}^\alpha + \left[\left(u+h\right)^\frac{\alpha}{\gamma}-h^\frac{\alpha}{\gamma}\right] \mathcal{E}_{AC}^\alpha,
\end{eqnarray*}
where $\frac{\mathcal{E}_{AB}^\gamma}{\mathcal{E}_{AC}^\gamma}\leq h, \frac{\alpha}{\gamma} \in[0,1]$, thus $0 \leq \alpha \leq \gamma$.
Moreover, if $\mathcal{E}_{AC}=0$, then $\mathcal{E}_{AB}=\mathcal{E}_{AC}=0$. This means that the lower bound of the inequality becomes zero. If $h\mathcal{E}_{AB}^\gamma \geq \mathcal{E}_{AC}^\gamma$, the inequality (\ref{p17}) can be obtained by a similar proof.

Here, we propose a general monogamous relation that holds for any entanglement measure and real number $\gamma$ satisfying the inequality (\ref{p1}). The new general monogamy relation can be applied to entanglement measures such as concurrence, entanglement of formation and the convex-roof extended negativity. The corresponding monogamous relations are better than the existing ones in \cite{s33,s35}, as well as complementary to the existing ones in \cite{s26}. We note that the third system $C$ in Theorem 1 can be divided into two subsystems: a qubit system $ C_1 $ and a $ 2^{n-3} $-dimensional system $ C_2 $. So, by repeating Theorem 1, we can generalize the monogamy inequality to multipartite qubit systems, namely Theorem 2.

{[\bf Theorem 2]}.
For any $n$-qubit state $\rho_{AB_1\cdots B_{n-1}} \in H_A\otimes H_{B_1}\otimes\cdots\otimes H_{B_{n-1}}$, let $u_p\geq1$ and $0\leq h_p\leq1$ be real numbers, $1\leq p\leq n-2$, if $\mathcal{E}^\gamma_{AB_i}\leq h_i\mathcal{E}^\gamma_{A|B_{i+1}\cdots B_{n-1}}$, $\mathcal{E}^\gamma_{A|B_{i}\cdots B_{n-1}}\geq \mathcal{E}^\gamma_{AB_i}+u_i\mathcal{E}^\gamma_{A|B_{i+1}\cdots B_{n-1}}$ for $i=1,2,\cdots ,m$, and $h_j\mathcal{E}^\gamma_{AB_j}\geq \mathcal{E}^\gamma_{A|B_{j+1}\cdots B_{n-1}}$, $\mathcal{E}^\gamma_{A|B_{j}\cdots B_{n-1}}\geq u_j\mathcal{E}^\gamma_{AB_j}+\mathcal{E}^\gamma_{A|B_{j+1}\cdots B_{n-1}}$ for $j=m+1,\cdots,{n-2}$, $1\leq m\leq {n-3},~n\geq 4$, then the entanglement measure $\mathcal{E}$ satisfies
\begin{eqnarray}\label{l6}
	\mathcal{E}^\alpha_{A|B_1B_2\cdots B_{n-1}}&&\geq \mathcal{E}^\alpha_{AB_1}
	+\varGamma_1 \mathcal{E}^\alpha_{AB_2}+\cdots+\varGamma_1\cdots\varGamma_{m-1}\mathcal{E}^\alpha_{AB_m} \nonumber\\
	&&\quad+\varGamma_1\cdots\varGamma_{m}\left(\varGamma_{m+1}\mathcal{E}^\alpha_{AB_{m+1}}
	+\cdots+\varGamma_{n-2}\mathcal{E}^\alpha_{AB_{n-2}}\right) \nonumber\\
	&&\quad+\varGamma_1\cdots\varGamma_{m}\mathcal{E}^\alpha_{AB_{n-1}}
\end{eqnarray}
for $0 \leq \alpha \leq \gamma$ and $\gamma \in Q$, where $\varGamma_p=\left(u_p+h_p\right)^\frac{\alpha}{\gamma}-h_p^\frac{\alpha}{\gamma}$.

{\sf [Proof].}
From Theorem 1, we have
\begin{eqnarray}\label{l7}
	\mathcal{E}^\alpha_{A|B_1B_2\cdots B_{n-1}}&&\geq \mathcal{E}_{AB_1}^\alpha + \varGamma_1 \mathcal{E}_{A|B_2\cdots B_{n-1}}^\alpha \nonumber \\
	&&\geq \mathcal{E}_{AB_1}^\alpha + \varGamma_1\mathcal{E}_{AB_2}^\alpha+ \varGamma_1\varGamma_2 \mathcal{E}_{A|B_3\cdots B_{n-1}}^\alpha \nonumber \\
	&&\geq \cdots \nonumber \\
	&&\geq \mathcal{E}_{AB_1}^\alpha + \varGamma_1 \mathcal{E}_{AB_2}^\alpha + \cdots + \varGamma_1\varGamma_2\cdots\varGamma_{m-1} \mathcal{E}_{AB_m}^\alpha \nonumber \\
	&&\quad+\varGamma_1\varGamma_2\cdots\varGamma_m \mathcal{E}_{A|B_{m+1}\cdots B_{n-1}}^\alpha.
\end{eqnarray}
Similarly, as $h_j\mathcal{E}^\gamma_{AB_j}\geq \mathcal{E}^\gamma_{A|B_{j+1}\cdots B_{n-1}}$ and $\mathcal{E}^\gamma_{A|B_{j}\cdots B_{n-1}}\geq u_j\mathcal{E}^\gamma_{AB_j}+\mathcal{E}^\gamma_{A|B_{j+1}\cdots B_{n-1}}$ for $j=m+1,\cdots,n-2$, we can get
\begin{eqnarray}\label{l8}
	\mathcal{E}_{A|B_{m+1}\cdots B_{n-1}}^\alpha
	&&\geq \varGamma_{m+1} \mathcal{E}_{AB_{m+1}}^\alpha + \mathcal{E}_{A|B_{m+2}\cdots B_{n-1}}^\alpha \nonumber \\
	&&\geq \varGamma_{m+1} \mathcal{E}_{A|B_{m+1}}^\alpha +\varGamma_{m+2} \mathcal{E}_{AB_{m+2}}^\alpha+ \mathcal{E}_{A|B_{m+3}\cdots B_{n-1}}^\alpha \nonumber \\
	&&\geq \cdots \nonumber \\
	&&\geq \varGamma_{m+1}\mathcal{E}^\alpha_{AB_{m+1}}+ \varGamma_{m+2}\mathcal{E}^\alpha_{AB_{m+2}}
	+\cdots+\varGamma_{n-2}\mathcal{E}^\alpha_{AB_{n-2}}  + \mathcal{E}_{AB_{n-1}}^\alpha.
\end{eqnarray}
Combining inequality (\ref{l7}) and (\ref{l8}), we have Theorem 1.

For the same entanglement measure and $\gamma$, our monogamy relation has a lower bound that is larger than the corresponding inequality in \cite{s33,s35}. Moreover, the relation is reduced to the existing ones given in \cite{s33,s35} if $ u  $ and $ h $ are given some specific values. When $m=n-2$, then the conditions in Theorem 2 are simplified to $\mathcal{E}^\gamma_{AB_i}\leq h_i\mathcal{E}^\gamma_{A|B_{i+1}\cdots B_{n-1}}$ and $\mathcal{E}^\gamma_{A|B_{i}\cdots B_{n-1}}\geq \mathcal{E}^\gamma_{AB_i}+u_i\mathcal{E}^\gamma_{A|B_{i+1}\cdots B_{n-1}}$ for all $i=1,2,\cdots ,n-2$. This leads to a special case of Theorem 2, namely Theorem 3.

{\bf [Theorem 3]}.
For any $n$-qubit state $\rho_{AB_1\cdots B_{n-1}} \in H_A\otimes H_{B_1}\otimes\cdots\otimes H_{B_{n-1}}$, let $u_p\geq1$ and $0\leq h_p\leq1$ be real numbers, $1\leq p\leq n-2$, if $\mathcal{E}^\gamma_{AB_i}\leq h_i\mathcal{E}^\gamma_{A|B_{i+1}\cdots B_{n-1}}$, $\mathcal{E}^\gamma_{A|B_{i}\cdots B_{n-1}}\geq \mathcal{E}^\gamma_{AB_i}+u_i\mathcal{E}^\gamma_{A|B_{i+1}\cdots B_{n-1}}$ for all $i=1,2,\cdots ,n-2$, then the entanglement measure $\mathcal{E}$ satisfies
\begin{eqnarray}
	\mathcal{E}^\alpha_{A|B_1B_2\cdots B_{n-1}} \geq \mathcal{E}^\alpha_{AB_1}
	+\varGamma_1 \mathcal{E}^\alpha_{AB_2}+\cdots+\varGamma_1\cdots\varGamma_{n-2}\mathcal{E}^\alpha_{AB_{n-1}}
\end{eqnarray}
for $0 \leq \alpha \leq \gamma$ and $\gamma \in Q$, where $\varGamma_p=\left(u_p+h_p\right)^\frac{\alpha}{\gamma}-h_p^\frac{\alpha}{\gamma}$.

We then apply Theorem 1 to concurrence, taking concurrence as an example to illustrate the advantages of our new results. It has been proved that concurrence satisfies $C_{A|BC}^\eta \geq C_{AB}^\eta + C_{AC}^\eta$ for any $2\otimes2\otimes2^{n-2}$ mixed tripartite state $\rho_{A|BC}$ \cite{s17} with $\eta \geq 2$. Thus, the following corollary can be obtained.

{\bf [Corollary 1]}.
For any $2\otimes2\otimes2^{n-2}$ mixed tripartite state $\rho_{ABC}\in H_A\otimes H_B\otimes H_C$, let real number $h$ satisfy $0\leq h\leq 1$, there exists real number $u\geq 1$,

(1) if $hC_{AC}^\gamma \geq C_{AB}^\gamma $, the concurrence satisfies
\begin{eqnarray}
	C_{A|BC}^\alpha \geq C_{AB}^\alpha + \left[\left(u+h\right)^\frac{\alpha}{\gamma}-h^\frac{\alpha}{\gamma}\right] C_{AC}^\alpha
\end{eqnarray}
for $0 \leq \alpha \leq \gamma$ and $\gamma \geq 2$.

(2) if $hC_{AB}^\gamma \geq C_{AC}^\gamma$, the concurrence satisfies
\begin{eqnarray}
	C_{A|BC}^\alpha \geq  \left[\left(u+h\right)^\frac{\alpha}{\gamma}-h^\frac{\alpha}{\gamma}\right] C_{AB}^\alpha + C_{AC}^\alpha
\end{eqnarray}
for $0 \leq \alpha \leq \gamma$ and $\gamma \geq 2$.

For $0\leq \alpha\leq \gamma$ with $\gamma\geq 2$, $u\geq 1$ and $0\leq h\leq 1$, we have
\begin{eqnarray}\label{l13}
	(u+h)^\frac{\alpha}{\gamma}-h^\frac{\alpha}{\gamma} \geq (1+h)^\frac{\alpha}{\gamma}-h^\frac{\alpha}{\gamma}\geq 2^\frac{\alpha}{\gamma}-1,
\end{eqnarray}
where the first and second equality hold when $u=1$ and $h=1$, respectively. From inequality (\ref{l13}), we can get $(1+h)^\frac{\alpha}{\gamma}-h^\frac{\alpha}{\gamma}=\frac{(1+k)^{\frac{\alpha}{\gamma}}-1}{k^{\frac{\alpha}{\gamma}}}$ if $h=\frac{1}{k}$ with $k\geq 1$, as well as $\frac{(1+k)^{\frac{\alpha}{\gamma}}-1}{k^{\frac{\alpha}{\gamma}}}\geq 2^\frac{\alpha}{\gamma}-1$ for $k \geq 1$ and $\frac{\alpha}{\gamma}\in [0,1]$. For a certain $h$, the larger $u$ is, the tighter the inequality in Corollary 1 is. Obviously, the new monogamy inequality of concurrence is better than the inequality in Ref. \cite{s35} if $h=\frac{1}{k}$ and $u>1$. In addition, Corollary 1 is reduced to the result in Ref. \cite{s35} if $h=\frac{1}{k}$ and $u=1$, as well as reduce to the result in Ref. \cite{s33} if $k=1$ and $u=1$. Therefore, our result is also tighter than the inequality in Ref. \cite{s33}.

\begin{figure}[h]
	\centering
	\scalebox{2.0}{\includegraphics[width=3.9cm]{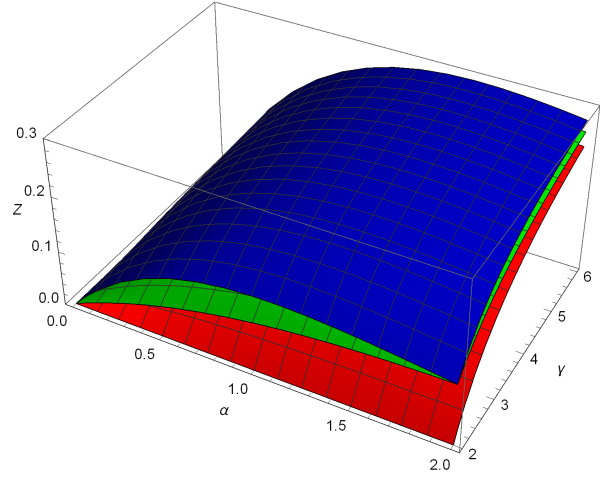}}
	\caption{$z$ is the ``residual" entanglement as a function of $\alpha, \gamma$. The blue surface represents the ``residual" entanglement given by $z_1=C^\alpha_{A|BC}-	C_{AB}^\alpha-(2^\frac{\alpha}{\gamma}-1) C_{AC}^\alpha$, the green surface represents the ``residual" entanglement given by $z_2=C^\alpha_{A|BC}-C_{AB}^\alpha - (\frac{(1+k)^{\frac{\alpha}{\gamma}}-1}{k^{\frac{\alpha}{\gamma}}}) C_{AC}^\alpha$, and the red surface represents the ``residual" entanglement given by $z_3=C^\alpha_{A|BC}-	C_{AB}^\alpha-[\left(u+h\right)^\frac{\alpha}{\gamma}-h^\frac{\alpha}{\gamma}] C_{AC}^\alpha$.}
	\label{fig1}
\end{figure}

{\bf Example 1}. Let us consider the three-qubit state $|\psi\rangle$, which can be written as \cite{s38}
\begin{eqnarray}\label{l10}
	|\psi\rangle=\lambda_0|000\rangle+\lambda_1e^{i\varphi}|100\rangle+\lambda_2|101\rangle+\lambda_3|110\rangle+\lambda_4|111\rangle,
\end{eqnarray}
where $\lambda_i \geq 0,i=0,\cdots,4$, and $\sum_{i=0}^4\lambda_i^2=1$. From the definition of concurrence, we have $C_{A|BC}=2\lambda_0\sqrt{\lambda_2^2+\lambda_3^2+\lambda_4^2}$, $C_{AB}=2\lambda_0\lambda_2$, and $C_{AC}=2\lambda_0\lambda_3$.
Set $\lambda_0=\lambda_1=\lambda_2=\lambda_4=\frac{1}{\sqrt{6}}, \lambda_3=\frac{\sqrt{3}}{3}$, one has $C_{A|BC}=\frac{2}{3}, C_{AB}=\frac{1}{3}, C_{AC}=\frac{\sqrt{2}}{3}$. Therefore,
\begin{eqnarray}\label{ll15}
	(\frac{2}{3})^\alpha \geq (\frac{1}{3})^\alpha +(2^\frac{\alpha}{\gamma}-1)(\frac{\sqrt{2}}{3})^\alpha,
\end{eqnarray}
\begin{eqnarray}\label{ll16}
	(\frac{2}{3})^\alpha \geq (\frac{1}{3})^\alpha + (\frac{(1+k)^{\frac{\alpha}{\gamma}}-1}{k^{\frac{\alpha}{\gamma}}}) (\frac{\sqrt{2}}{3})^\alpha,
\end{eqnarray}
\begin{eqnarray}\label{ll17}
	(\frac{2}{3})^\alpha \geq (\frac{1}{3})^\alpha + [\left(u+h\right)^\frac{\alpha}{\gamma}-h^\frac{\alpha}{\gamma}] (\frac{\sqrt{2}}{3})^\alpha.
\end{eqnarray}
When $k=2$, the lower bound of inequality (\ref{ll16}) gives the best result for $1\leq k\leq 2$ and $\frac{\alpha}{\gamma}\in [0,1]$. When $h=\frac{1}{k}$ and $u=1$, inequality (\ref{ll17}) is reduced to inequality (\ref{ll16}). However, when $1\geq h\geq \frac{1}{2}, \frac{3}{2} \geq u > 1$, the lower bound of inequality (\ref{ll17}) is better than that of inequality (\ref{ll16}). Obviously, let $h=\frac{1}{2}, u=\frac{3}{2}, k=\frac{1}{h}$, our result (\ref{ll17}) is better than the result (\ref{ll16}) given in Ref. \cite{s35} and the inequality (\ref{ll15}) given in Ref. \cite{s33}, see Fig \ref{fig1}.

Again, we use concurrence as an example to demonstrate the advantages of monogamy relation for multipartite systems. In Theorem 2 and Theorem 3, set $\mathcal{E}$ is concurrence, we will have the following two corollaries.

{\bf [Corollary 2]}.
For any $n$-qubit state $\rho_{AB_1\cdots B_{n-1}} \in H_A\otimes H_{B_1}\otimes\cdots\otimes H_{B_{n-1}}$, let $u_p\geq1$ and $0\leq h_p\leq1$ be real numbers, $1\leq p\leq n-2$, if $C^\gamma_{AB_i}\leq h_iC^\gamma_{A|B_{i+1}\cdots B_{n-1}}$, $C^\gamma_{A|B_{i}\cdots B_{n-1}}\geq C^\gamma_{AB_i}+u_iC^\gamma_{A|B_{i+1}\cdots B_{n-1}}$ for $i=1,2,\cdots ,m$, and $h_jC^\gamma_{AB_j}\geq C^\gamma_{A|B_{j+1}\cdots B_{n-1}}$, $C^\gamma_{A|B_{j}\cdots B_{n-1}}\geq u_jC^\gamma_{AB_j}+C^\gamma_{A|B_{j+1}\cdots B_{n-1}}$ for $j=m+1,\cdots,{n-2}$, $1\leq m\leq {n-3},~n\geq 4$, then the concurrence satisfies
\begin{eqnarray}\label{l18}
	C^\alpha_{A|B_1B_2\cdots B_{n-1}}&&\geq C^\alpha_{AB_1}
	+\varGamma_1 C^\alpha_{AB_2}+\cdots+\varGamma_1\cdots\varGamma_{m-1}C^\alpha_{AB_m} \nonumber\\
	&&\quad+\varGamma_1\cdots\varGamma_{m}\left(\varGamma_{m+1}C^\alpha_{AB_{m+1}}
	+\cdots+\varGamma_{n-2}C^\alpha_{AB_{n-2}}\right) \nonumber\\
	&&\quad+\varGamma_1\cdots\varGamma_{m}C^\alpha_{AB_{n-1}}
\end{eqnarray}
for $0 \leq \alpha \leq \gamma$ and $\gamma \geq 2$, where $\varGamma_p=\left(u_p+h_p\right)^\frac{\alpha}{\gamma}-h_p^\frac{\alpha}{\gamma}$.

{\bf [Corollary 3]}.
For any $n$-qubit state $\rho_{AB_1\cdots B_{n-1}} \in H_A\otimes H_{B_1}\otimes\cdots\otimes H_{B_{n-1}}$, let $u_p\geq1$ and $0\leq h_p\leq1$ be real numbers, $1\leq p\leq n-2$, if $C^\gamma_{AB_i}\leq h_iC^\gamma_{A|B_{i+1}\cdots B_{n-1}}$, $C^\gamma_{A|B_{i}\cdots B_{n-1}}\geq C^\gamma_{AB_i}+u_iC^\gamma_{A|B_{i+1}\cdots B_{n-1}}$ for all $i=1,2,\cdots ,n-2$, then the concurrence satisfies
\begin{eqnarray}\label{l6}
	C^\alpha_{A|B_1B_2\cdots B_{n-1}} \geq C^\alpha_{AB_1}
	+\varGamma_1 C^\alpha_{AB_2}+\cdots+\varGamma_1\cdots\varGamma_{n-2}C^\alpha_{AB_{n-1}}
\end{eqnarray}
for $0 \leq \alpha \leq \gamma$ and $\gamma \geq 2$, where $\varGamma_p=\left(u_p+h_p\right)^\frac{\alpha}{\gamma}-h_p^\frac{\alpha}{\gamma}$.

Considering the case of $\gamma=2$, we can see that the inequalities in Corollary 2 and Corollary 3 are complementary to the inequalities in Ref. \cite{s26}. If $h_p=1$ and $u_p=1$ for all $p$, the inequalities in Corollary 1 and Corollary 2 reduce to the monogamous relation in Ref. \cite{s33}. Otherwise, our results are better than the existing ones in Ref. \cite{s33}. Our inequality in Corollary 2 is also tighter than the result in \cite{s35} for all $h_p=\frac{1}{k}$ and $u_p>1$, as well as reduce to the relation in Ref. \cite{s35} if $h_p=\frac{1}{k}$ and $u_p=1$ for all $p$.
In particular, the negativity has a monogamy relation similar to Corollary 2 for pure states satisfying the same conditions. The new monogamy relation will better than the following ones.

For any $N$-qubit pure state $|\psi\rangle_{A|B_1\cdots B_{N-1}}$, it has been shown that the negativity satisfies \cite{s21}
\begin{eqnarray}
	[N(|\psi\rangle_{A|B_1,B_2\cdots,B_{n-1}})]^\alpha\geq \sum_{i=1}^{n-1}N_{AB_i}^\alpha
\end{eqnarray}
for $\alpha \geq 2$. It has been further proved that for any $n$-qubit quantum pure state $|\psi\rangle_{A|B_1\cdots B_{n-1}}$, the negativity satisfies the monogamy relation for $0\leq \alpha \leq \gamma$ and $\gamma \geq 2$ \cite{s33}:
\begin{eqnarray}\label{l21}
	[N(|\psi\rangle_{A|B_1,B_2\cdots,B_{n-1}})]^\alpha\geq \sum_{i=1}^m(2^\frac{\alpha}{\gamma}-1)^{i-1}N^\alpha_{AB_i}+(2^\frac{\alpha}{\gamma}-1)^{m+1}\sum_{i=m+1}^{n-2}N^\alpha_{AB_i}+(2^\frac{\alpha}{\gamma}-1)^mN^\alpha_{AB_{n-1}},
\end{eqnarray}
where $C_{AB_i}\leq C_{A|B_{i+1}\cdots B_{n-1}}$ for $i=1,\cdots,m$, and $C_{AB_j}\geq C_{A|B_{j+1}\cdots B_{n-1}}$ for $j=m+1,\cdots,n-2$, $1\leq m \leq n-3$, $n\geq 4$.

 For $2\otimes v \otimes w$ systems, one has $C(|\psi\rangle_{A|BC})=N(|\psi\rangle_{A|BC})$ with $v \geq2$ and $w\geq 2$  \cite{s21}. By extending this equation to multipartite qubit systems, we get $N(|\psi\rangle_{A|B_1,B_2\cdots,B_{n-1}})=C(|\psi\rangle_{A|B_1,B_2\cdots,B_{n-1}}) $. Moreover, inequality $N(\rho_{AB})\leq C(\rho_{AB})$ holds for $2\otimes d$ systems \cite{s18}. Therefore,
  using Corollary 2 and the relation between concurrence and negativity, we can obtain a similar result for the negativity of pure states.

{\bf [Corollary 4]}.
For any $n$-qubit pure state $|\psi\rangle_{A|B_1\cdots B_{n-1}}$, let $u_p \geq 1, 0\leq h_p\leq1$, if $C^\gamma_{AB_i}\leq h_iC^\gamma_{A|B_{i+1}\cdots B_{n-1}}$, $C^\gamma_{A|B_{i}\cdots B_{n-1}}\geq C^\gamma_{AB_i}+u_iC^\gamma_{A|B_{i+1}\cdots B_{n-1}}$ for $i=1,2,\cdots ,m$, and $h_jC^\gamma_{AB_j}\geq C^\gamma_{A|B_{j+1}\cdots B_{n-1}}$, $C^\gamma_{A|B_{j}\cdots B_{n-1}}\geq u_jC^\gamma_{AB_j}+C^\gamma_{A|B_{j+1}\cdots B_{n-1}}$ for $j=m+1,\cdots,{n-2}$, $1\leq m\leq {n-3},~n\geq 4$, then the negativity satisfies
\begin{eqnarray}\label{p22}
	 [N(|\psi\rangle_{A|B_1,B_2\cdots,B_{n-1}})]^\alpha
	 &&\geq N^\alpha_{AB_1}
	+\varGamma_1 N^\alpha_{AB_2}+\cdots+\varGamma_1\cdots\varGamma_{m-1}N^\alpha_{AB_m} \nonumber\\
	&&\quad+\varGamma_1\cdots\varGamma_{m}\left(\varGamma_{m+1}N^\alpha_{AB_{m+1}}
	+\cdots+\varGamma_{n-2}N^\alpha_{AB_{n-2}}\right) \nonumber\\
	&&\quad+\varGamma_1\cdots\varGamma_{m}N^\alpha_{AB_{n-1}}
\end{eqnarray}
for $0 \leq \alpha \leq \gamma$ and $\gamma \geq 2$, where $\varGamma_p=\left(u_p+h_p\right)^\frac{\alpha}{\gamma}-h_p^\frac{\alpha}{\gamma}$.

From the previous analysis of Corollary 1, it can be seen that our result is better than the inequality (\ref{l21}) from Ref. \cite{s33}. Obviously, as a result of $N(|\psi\rangle_{A|B_1,B_2\cdots,B_{n-1}})=C(|\psi\rangle_{A|B_1,B_2\cdots,B_{n-1}})$, so the lower bound of inequality (\ref{p22}) is also the lower bound of $C(|\psi\rangle_{A|B_1,B_2\cdots,B_{n-1}})$ for $n$-qubit pure state.

\section{general Polygamy properties for assisted entanglement measures}
Let $\mathcal{E}_a$ be a quantum assisted entanglement measure. For any $2\otimes2\otimes2^{n-2}$ tripartite state $\rho_{A|BC} \in H_A\otimes H_B\otimes H_C$, assume non-negative real number $\delta$ is the value for $\mathcal{E}_a^\delta$ to satisfy inequality
\begin{eqnarray}\label{l23}
	\mathcal{E}_{aA|BC}^\delta \leq \mathcal{E}_{aAB}^\delta +  \mathcal{E}_{aAC}^\delta,
\end{eqnarray}
that is, $\delta \in Q' = \{\eta ~|~\mathcal{E}_{aA|BC}^\eta \leq \mathcal{E}_{aAB}^\eta +  \mathcal{E}_{aAC}^\eta$ for all  tripartite state $ \rho_{ABC} \}$.
It has been shown that assisted entanglement measures, such as concurrence of assistance \cite{s28}, entanglement of assistance \cite{s29} and the convex-roof extended negativity of assistance \cite{s39}, satisfy the polygamy relation (\ref{l23}).

For the $\beta$th power of assisted entanglement measure $\mathcal{E}_a^\beta$, we will propose a new class of polygamy relation for multipartite quantum states with some constraints according to Lemma 1, which are tighter than the inequalities in Ref. \cite{s35}.

{[\bf Theorem 4]}.
For any $2\otimes2\otimes2^{n-2}$ tripartite state $\rho_{A|BC}\in H_A\otimes H_B\otimes H_C$, let real number $h$ satisfy $0\leq h\leq 1$, there exists real number $0\leq u\leq 1$:

(1) if $\mathcal{E}_{aAB}^\delta \leq h\mathcal{E}_{aAC}^\delta$, the assisted entanglement satisfies
\begin{eqnarray}\label{p13}
	\mathcal{E}_{aA|BC}^\beta \leq \mathcal{E}_{aAB}^\beta + \left[\left(u+h\right)^\frac{\beta}{\delta}-h^\frac{\beta}{\delta}\right] \mathcal{E}_{aAC}^\beta
\end{eqnarray}
for $\beta \geq \delta$ and $\delta \in Q'$.

(2) if $h\mathcal{E}_{aAB}^\delta \geq \mathcal{E}_{aAC}^\delta$, the assisted entanglement satisfies
\begin{eqnarray}\label{p12}
	\mathcal{E}_{aA|BC}^\beta \leq  \left[\left(u+h\right)^\frac{\beta}{\delta}-h^\frac{\beta}{\delta}\right] \mathcal{E}_{aAB}^\beta + \mathcal{E}_{aAC}^\beta
\end{eqnarray}
for $\beta \geq \delta$ and $\delta \in Q'$.

{\sf [Proof].}
For any $2\otimes2\otimes2^{n-2}$ quantum state $\rho_{ABC}\in H_A\otimes H_B\otimes H_C$, one has relation $\mathcal{E}_{aA|BC}^\delta \leq \mathcal{E}_{aAB}^\delta + \mathcal{E}_{aAC}^\delta$.
Therefore, there exists $0\leq u\leq 1$ such that
\begin{eqnarray}\label{p15}
    \mathcal{E}_{aA|BC}^\delta \leq \mathcal{E}_{aAB}^\delta + u\mathcal{E}_{aAC}^\delta.
\end{eqnarray}

Using the relation (\ref{p15}) and inequality (\ref{p3}) in Lemma 1, we have
\begin{eqnarray*}\label{p10}
	\mathcal{E}_{aA|BC}^\beta
	&&\leq (\mathcal{E}_{aAB}^\delta + u\mathcal{E}_{aAC}^\delta)^\frac{\beta}{\delta}\\
	&& = \mathcal{E}_{aAB}^\beta +u^\frac{\beta}{\delta}\mathcal{E}_{aAC}^\beta\left[\left(\frac{1}{u}\frac{\mathcal{E}_{aAB}^\delta}{\mathcal{E}_{aAC}^\delta}+1\right)^\frac{\beta}{\delta}-\left(\frac{1}{u}\frac{\mathcal{E}_{aAB}^\delta}{\mathcal{E}_{aAC}^\delta}\right)^\frac{\beta}{\delta}\right]\\
	&&\leq \mathcal{E}_{aAB}^\beta +u^\frac{\beta}{\delta}\mathcal{E}_{aAC}^\beta\left[\left(\frac{h}{u}+1\right)^\frac{\beta}{\delta}-\left(\frac{h}{u}\right)^\frac{\beta}{\delta}\right] \\
	&& = \mathcal{E}_{aAB}^\beta + \left[\left(u+h\right)^\frac{\beta}{\delta}-h^\frac{\beta}{\delta}\right] \mathcal{E}_{aAC}^\beta,
\end{eqnarray*}
where $\frac{\mathcal{E}_{aAB}^\delta}{\mathcal{E}_{aAC}^\delta}\leq h,~ \frac{\beta}{\delta}\geq 1$, thus $\beta \geq \delta$.
Moreover, if $\mathcal{E}_{aAC}=0$, then $\mathcal{E}_{aAB}=\mathcal{E}_{aAC}=0$. This means that the lower bound of the inequality becomes zero. If $h\mathcal{E}_{aAB}^\delta \geq \mathcal{E}_{aAC}^\delta$, the inequality (\ref{p12}) can be obtained by a similar proof.

We establish a general polygamous relation for any assisted entanglement measure and real number $\delta$ satisfying inequality (\ref{l23}). The new general polygamy relation can be applied to any assisted entanglement measures like concurrence of assistance, entanglement of assistance and the convex-roof extended negativity of assistance. The corresponding polygamy relations are tighter than the existing ones in \cite{s35}. In the same way, the third system $C$ in Theorem 4 can be divided into a qubit system $ C_1 $ and a $ 2^{n-3} $-dimensional system $ C_2 $, so we can generalize the polygamy inequality to multipartite qubit systems by using Theorem 4 repeatedly. Then we have Theorem 5.

{\bf [Theorem 5]}.
For any $n$-qubit quantum state $\rho_{A|B_1\cdots B_{n-1}} \in H_A\otimes H_{B_1}\otimes\cdots\otimes H_{B_{n-1}}$, let $0\leq u_p\leq1$ and $0\leq h_p\leq1$ be real numbers, $1\leq p\leq n-2$, if $\mathcal{E}^\delta_{aAB_i} \leq h_i\mathcal{E}^\delta_{aA|B_{i+1}\cdots B_{n-1}}$, $\mathcal{E}^\delta_{aA|B_{i}\cdots B_{n-1}}\leq \mathcal{E}^\delta_{aAB_i}+u_i\mathcal{E}^\delta_{aA|B_{i+1}\cdots B_{n-1}}$ for $i=1,2,\cdots ,m$, and $h_j\mathcal{E}^\delta_{aAB_j}\geq \mathcal{E}^\delta_{aA|B_{j+1}\cdots B_{n-1}}$, $\mathcal{E}^\delta_{aA|B_{j}\cdots B_{n-1}}\leq u_j\mathcal{E}^\delta_{aAB_j}+\mathcal{E}^\delta_{aA|B_{j+1}\cdots B_{n-1}}$ for $j=m+1,\cdots,{n-2}$, $1\leq m\leq {n-3},~n\geq 4$, then the assisted entanglement satisfies
\begin{eqnarray}
	\mathcal{E}_{aA|B_1B_2\cdots B_{n-1}}^\beta &&\leq \mathcal{E}^\beta_{aAB_1}
	+\varPi_1 \mathcal{E}^\beta_{aAB_2}+\cdots+\varPi_1\cdots\varPi_{m-1}\mathcal{E}^\beta_{aAB_m} \nonumber\\
	&&\quad+\varPi_1\cdots\varPi_{m}\left(\varPi_{m+1}\mathcal{E}^\beta_{aAB_{m+1}}
	+\cdots+\varPi_{n-2}\mathcal{E}^\beta_{aAB_{n-2}}\right) \nonumber\\
	&&\quad+\varPi_1\cdots\varPi_{m}\mathcal{E}^\beta_{aAB_{n-1}}
\end{eqnarray}
for $\beta \geq \delta$ and $\delta \in Q'$, where $\varPi_p=\left(u_p+h_p\right)^\frac{\beta}{\delta}-h_p^\frac{\beta}{\delta}$.

{\sf [Proof].}
By using Theorem 4 repeatedly, the proof process is similar to Theorem 2.

For the same assisted entanglement measure, the upper bound of the new polygamy relation is smaller than that of the corresponding inequality in Ref. \cite{s35}. If we are going to the conditions in Theorem 5 is simplified to $\mathcal{E}^\delta_{aAB_i}\leq h_i\mathcal{E}^\delta_{aA|B_{i+1}\cdots B_{n-1}}$, $\mathcal{E}^\delta_{aA|B_{i}\cdots B_{n-1}}\leq \mathcal{E}^\delta_{aAB_i}+u_i\mathcal{E}^\delta_{aA|B_{i+1}\cdots B_{n-1}}$ for $i=1,2,\cdots ,n-2$, a special case of Theorem 5 will be obtained, which is Theorem 6.

{\bf [Theorem 6]}.
For any $n$-qubit quantum state $\rho_{A|B_1\cdots B_{n-1}} \in H_A\otimes H_{B_1}\otimes\cdots\otimes H_{B_{n-1}}$, let $0\leq u_p\leq1$ and $0\leq h_p\leq1$ be real numbers, $1\leq p\leq n-2$, if $\mathcal{E}^\delta_{aAB_i}\leq h_i\mathcal{E}^\delta_{aA|B_{i+1}\cdots B_{n-1}}$, $\mathcal{E}^\delta_{aA|B_{i}\cdots B_{n-1}}\leq \mathcal{E}^\delta_{aAB_i}+u_i\mathcal{E}^\delta_{aA|B_{i+1}\cdots B_{n-1}}$ for $i=1,2,\cdots ,n-2$, then the assisted entanglement satisfies
\begin{eqnarray}
	\mathcal{E}_{aA|B_1B_2\cdots B_{n-1}}^\beta \leq \mathcal{E}^\beta_{aAB_1}
	+\varPi_1 \mathcal{E}^\beta_{aAB_2}+\cdots+\varPi_1\cdots\varPi_{n-2}\mathcal{E}^\beta_{aAB_{n-1}}
\end{eqnarray}
for $\beta \geq \delta$ and $\delta \in Q'$, where $\varPi_p=\left(u_p+h_p\right)^\frac{\beta}{\delta}-h_p^\frac{\beta}{\delta}$.

We present a general form of polygamy relation for multipartite system that is better than the corresponding relation in Ref. \cite{s35}. For a specific measure like entanglement of formation, it is complementary to the corresponding polygamy relation in Ref. \cite{s26}. Next, we apply Theorem 4 to CRENoA as an example to illustrate the advantages of the new relation. In Ref. \cite{s39}, the authors proved that $\widetilde{N}_{aA|BC}^\delta \leq \widetilde{N}_{aAB}^\delta + \widetilde{N}_{aAC}^\delta$ with $0 \leq \delta \leq 2$ for arbitrary $2\otimes2\otimes2^{n-2}$ tripartite states. Therefore, set the measure $\mathcal{E}_a $ in Theorem 4 is CRENoA, we can get

{\bf [Corollary 5]}.
For any $2\otimes2\otimes2^{n-2}$ quantum state $\rho_{A|BC}\in H_A\otimes H_B\otimes H_C$, set real number $h$ satisfy $0\leq h\leq 1$, there exists real number $0\leq u\leq 1$ such that

(1) if $\widetilde{N}_{aAB}^\delta \leq h\widetilde{N}_{aAC}^\delta$, the CRENoA satisfies
\begin{eqnarray}\label{p13}
	\widetilde{N}_{aA|BC}^\beta \leq \widetilde{N}_{aAB}^\beta + \left[\left(u+h\right)^\frac{\beta}{\delta}-h^\frac{\beta}{\delta}\right] \widetilde{N}_{aAC}^\beta
\end{eqnarray}
for $\beta \geq \delta$ and $0 \leq \delta \leq 2$.

(2) if $h\widetilde{N}_{aAB}^\delta \geq \widetilde{N}_{aAC}^\delta$, the CRENoA satisfies
\begin{eqnarray}\label{p14}
	\widetilde{N}_{aA|BC}^\beta \leq  \left[\left(u+h\right)^\frac{\beta}{\delta}-h^\frac{\beta}{\delta}\right] \widetilde{N}_{aAB}^\beta + \widetilde{N}_{aAC}^\beta
\end{eqnarray}
for $\beta \geq \delta$ and $0 \leq \delta \leq 2$.

For $\beta\geq \delta$ with $\delta \geq 1$, $0\leq h \leq 1$ and $0\leq u\leq 1$ , we have
\begin{eqnarray}\label{c13}
	(u+h)^\frac{\beta}{\delta}-h^\frac{\beta}{\delta} \leq (1+h)^\frac{\beta}{\delta}-h^\frac{\beta}{\delta} \leq 2^\frac{\beta}{\delta}-1,
\end{eqnarray}
where the first and second equality hold when $u=1$ and $h=1$, respectively. From inequality (\ref{c13}), we can obtain $(1+h)^\frac{\beta}{\delta}-h^\frac{\beta}{\delta}=\frac{(1+k)^{\frac{\beta}{\delta}}-1}{k^{\frac{\beta}{\delta}}}$ if $h=\frac{1}{k}$ with $k\geq 1$, as well as $\frac{(1+k)^{\frac{\beta}{\delta}}-1}{k^{\frac{\beta}{\delta}}}\leq2^\frac{\beta}{\delta}-1$ for $k \geq 1$ and $\frac{\beta}{\delta}\geq 1$.  For a certain $h$, the smaller $u$ is, the better the result in Corollary 4 is. Therefore, our new polygamy inequality of CRENoA is better than the existing ones in \cite{s35} if $h=\frac{1}{k}$ and $0\leq u < 1$, as well as Corollary 4 is reduced to the result in Ref. \cite{s35} if $h=\frac{1}{k}$ and $u=1$.

\begin{figure}[h]
	\centering
	\scalebox{2.0}{\includegraphics[width=3.9cm]{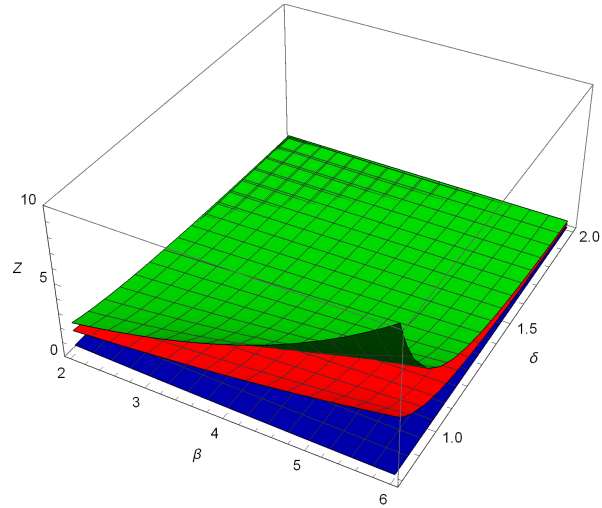}}
	\caption{The axis $z$ denotes the CRENoA and its upper bound, which are functions of $\beta$ and $\delta$. The blue surface represents the entanglement of three-qubit pure state, the green surface represents the upper bound from the inequality (\ref{b16}) given by \cite{s35}, and the red surface represents the upper bound from our result.}
	\label{fig2}
\end{figure}

{\bf Example 2}. Let us consider the same three-qubit pure state as in Example 1. By the calculation of CRENoA, we have \cite{s34} $\widetilde{N}_{aA|BC}=2\lambda_0\sqrt{\lambda_2^2+\lambda_3^2+\lambda_4^2}$, $\widetilde{N}_{aAB}=2\lambda_0\sqrt{\lambda^2_2+\lambda^2_4}$, and $\widetilde{N}_{aAC}=2\lambda_0\sqrt{\lambda^2_3+\lambda^2_4}$. Set $ \lambda_0=\lambda_4=\sqrt{\frac{2}{9}}, \lambda_1=\lambda_2= \frac{1}{3},\lambda_3=\sqrt{\frac{1}{3}}$, we have $\widetilde{N}_{aA|BC}=\frac{\sqrt{48}}{9}, \widetilde{N}_{aAB}=\frac{\sqrt{24}}{9}, \widetilde{N}_{aAC}=\frac{\sqrt{40}}{9}$. Therefore,
\begin{eqnarray}\label{b16}
	\left(\frac{\sqrt{48}}{9}\right)^\beta \leq  \left(\frac{\sqrt{24}}{9}\right)^\beta + (\frac{(1+k)^{\frac{\beta}{\delta}}-1}{k^{\frac{\beta}{\delta}}}) \left(\frac{\sqrt{40}}{9}\right)^\beta,
\end{eqnarray}
\begin{eqnarray}\label{b17}
	\left(\frac{\sqrt{48}}{9}\right)^\beta \leq  \left(\frac{\sqrt{24}}{9}\right)^\beta + \left[\left(u+h\right)^\frac{\beta}{\delta}-h^\frac{\beta}{\delta}\right] \left(\frac{\sqrt{40}}{9}\right)^\beta.
\end{eqnarray}
When $h=\frac{1}{k}$ and $u=1$, inequality (\ref{b17}) is reduced to inequality (\ref{b16}). For given $h\in[\frac{3}{5},1]$, the upper bound of (\ref{b17}) is better than that of (\ref{b16}) with arbitrary $u\in[\frac{3}{5},1)$. Obviously, let $h=\frac{4}{5}$, $u=\frac{4}{5}$ and $k=\frac{1}{h}$, our result is better than the results given in \cite{s35}, see Fig \ref{fig2}.

\begin{figure}[h]
	\centering
	\scalebox{2.0}{\includegraphics[width=3.9cm]{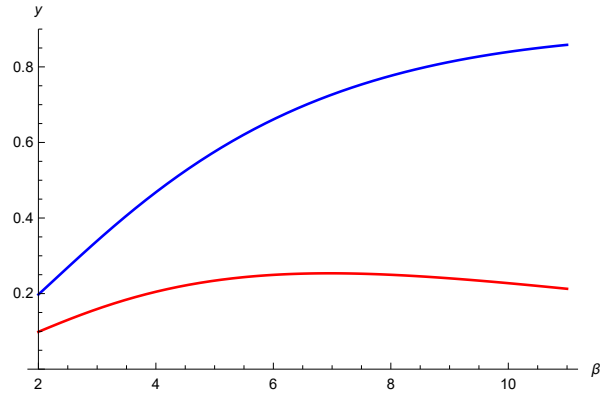}}
	\caption{The axis $y$ denotes the ``residual" entanglement, which are functions of $\beta$. The blue line represents the result in \cite{s34}, and the red line denotes our result. }
	\label{fig3}
\end{figure}

For the case of three-qubit system, the relation in Corollary 5 reduces to the Theorem 6 in Ref. \cite{s34} if $u=1,~h=1$ and $\delta=2$. Furthermore, the inequality in Corollary 5 tighter than the result in Ref. \cite{s34} if $0\leq u <1, ~ 0\leq h<1$ and $\delta=2$. Using the state in Example 2, set $u=\frac{4}{5},~h=\frac{4}{5}$ and $\delta=2$, we have $ y_1= \widetilde{N}_{aAB}^\beta + (2^\frac{\beta}{2}-1) \widetilde{N}_{aAC}^\beta -\widetilde{N}_{aA|BC}^\beta $  and
$y_2 =  \widetilde{N}_{aAB}^\beta +[ (\frac{8}{5})^\frac{\beta}{2}-(\frac{4}{5})^\frac{\beta}{2}] \widetilde{N}_{aAC}^\beta - \widetilde{N}_{aA|BC}^\beta $; see Fig 3.

Again, we use CRENoA as an example to demonstrate the advantages of polygamy relation for multipartite systems. If we set $\mathcal{E}_a$ in Theorem 5 and Theorem 6 is CRENoA, then we can obtain the following two corollaries.

{\bf [Corollary 6]}.
For any $n$-qubit quantum state $\rho_{A|B_1\cdots B_{n-1}} \in H_A\otimes H_{B_1}\otimes\cdots\otimes H_{B_{n-1}}$, set $0\leq u_p\leq1$ and $0\leq h_p\leq1$ be real numbers, $1\leq p\leq n-2$, if $\widetilde{N}^\delta_{aAB_i}\leq h_i\widetilde{N}^\delta_{aA|B_{i+1}\cdots B_{n-1}}$, $\widetilde{N}^\delta_{aA|B_{i}\cdots B_{n-1}}\leq \widetilde{N}^\delta_{aAB_i}+u_i\widetilde{N}^\delta_{aA|B_{i+1}\cdots B_{n-1}}$ for $i=1,2,\cdots ,m$, and $h_j\widetilde{N}^\delta_{aAB_j}\geq \widetilde{N}^\delta_{aA|B_{j+1}\cdots B_{n-1}}$, $\widetilde{N}^\delta_{aA|B_{j}\cdots B_{n-1}}\leq u_j\widetilde{N}^\delta_{aAB_j}+\widetilde{N}^\delta_{aA|B_{j+1}\cdots B_{n-1}}$ for $j=m+1,\cdots,{n-2}$, $1\leq m\leq {n-3},~n\geq 4$, then the CRENoA satisfies
\begin{eqnarray}
	\widetilde{N}_{aA|B_1\cdots B_{n-1}}^\beta \leq &&\widetilde{N}^\beta_{aAB_1}
	+\varPi_1 \widetilde{N}^\beta_{aAB_2}+\cdots+\varPi_1\cdots\varPi_{m-1}\widetilde{N}^\beta_{aAB_m} \nonumber\\
	&&+\varPi_1\cdots\varPi_{m}\left(\varPi_{m+1}\widetilde{N}^\beta_{aAB_{m+1}}
	+\cdots+\varPi_{n-2}\widetilde{N}^\beta_{aAB_{n-2}}\right) \nonumber\\
	&&+\varPi_1\cdots\varPi_{m}\widetilde{N}^\beta_{aAB_{n-1}}
\end{eqnarray}
for $\beta \geq \delta$ and $0 \leq \delta \leq 2$, where $\varPi_p=\left(u_p+h_p\right)^\frac{\beta}{\delta}-h_p^\frac{\beta}{\delta}$.

{\bf [Corollary 7]}.
For any $n$-qubit quantum state $\rho_{A|B_1\cdots B_{n-1}} \in H_A\otimes H_{B_1}\otimes\cdots\otimes H_{B_{n-1}}$, let $0\leq u_p\leq1$ and $0\leq h_p\leq1$ be real numbers, $1\leq p\leq n-2$, if $\widetilde{N}^\delta_{aAB_i}\leq h_i\widetilde{N}^\delta_{aA|B_{i+1}\cdots B_{n-1}}$, $\widetilde{N}^\delta_{aA|B_{i}\cdots B_{n-1}}\leq \widetilde{N}^\delta_{aAB_i}+u_i\widetilde{N}^\delta_{aA|B_{i+1}\cdots B_{n-1}}$ for $i=1,2,\cdots ,n-2$, then the CRENoA satisfies
\begin{eqnarray}\label{e35}
	\widetilde{N}_{aA|B_1\cdots B_{n-1}}^\beta \leq \widetilde{N}^\beta_{aAB_1}
	+\varPi_1 \widetilde{N}^\beta_{aAB_2}+\cdots+\varPi_1\cdots\varPi_{n-2}\widetilde{N}^\beta_{aAB_{n-1}}
\end{eqnarray}
for $\beta \geq \delta$ and $0 \leq \delta \leq 2$, where $\varPi_p=\left(u_p+h_p\right)^\frac{\beta}{\delta}-h_p^\frac{\beta}{\delta}$.

Now consider the case of $u_p=1$ and $h_p=\frac{1}{k}$ for all $p$, we get Corollary 5 and Corollary 6 reduce to the results in Ref. \cite{s35}. If $0\leq u_p<1$ and $h_p=\frac{1}{k}$ for all $p$, then the analysis of inequality (\ref{c13}) shows that the inequalities in Corollary 5 and Corollary 6 are tighter than the relations in Ref. \cite{s35}.

\section{conclusion}
Entanglement monogamy is an essential property of multipartite qubit systems and can characterize the distributions of entanglement. In this paper, we investigate general monogamy properties related to entanglement measures that satisfy the condition $\mathcal{E}_{A|BC}^\gamma\geq \mathcal{E}^\gamma_{AB}+\mathcal{E}^\gamma_{AC}$ for any $2\otimes2\otimes2^{n-2} $ state $\rho_{ABC}$. We also investigate polygamy properties related to any assisted entanglement measures. These new monogamy and polygamy relations describe the distribution of entanglement more precisely under stronger constraints. On the other side, our results complement the existing inequalities, which have different parameter regions. For a particular entanglement measure, the corresponding monogamous and polygamous relations take the existing ones as a special case. It is worth mentioning that other entanglement measures such as Tsallis-$q$ entanglement and R\'{e}nyi-$\alpha$ entanglement may have similar properties.

\bigskip
\noindent{\bf Acknowledgments}\, \, This work is supported by the NSF of China under Grant No. 12175147, 12171044.

\bigskip
\noindent{\bf Data availability statement}

All data generated or analysed during this study are included in this published article.

\bigskip
\noindent{\bf  Declarations}

\textbf{Conflict of interest}
The authors declare no competing interests.


\begin{thebibliography}{18}
\bibitem{s1} F. Mintert, M. Ku$\acute{s}$, and A. Buchleitner, Concurrence of Mixed
Bipartite Quantum States in Arbitrary Dimensions, Phys. Rev. Lett. 92, 167902 (2004).
\bibitem{s2} K. Chen, S. Albeverio, and S. M. Fei, Concurrence of arbitrary dimensional bipartite quantum states, Phys. Rev. Lett. 95, 040504 (2005).
\bibitem{s3}  H. P. Breuer, Separability criteria and bounds for entanglement measures, J. Phys. A: Math. Gen. 39, 11847 (2006).
\bibitem{s4} H. P. Breuer, Optimal entanglement criterion for mixed quantum states, Phys. Rev. Lett. 97, 080501 (2006).
\bibitem{s5} J. I. de Vicente, Lower bounds on concurrence and separability conditions, Phys. Rev. A 75, 052320 (2007).
\bibitem{s6} C. J. Zhang, Y. S. Zhang, S. Zhang, and G. C. Guo, Optimal entanglement witnesses based on local orthogonal observables, Phys. Rev. A 76, 012334 (2007).
\bibitem{s7} C. H. Bennett, G. Brassard, C. Cr$\acute{e}$peau, et al. Teleporting an unknown quantum state via dual classical and Einstein-Podolsky-Rosen channels, Phys. Rev. Lett. 70, 1895 (1993).
\bibitem{s8} M. Boyer, G. Ran, K. Dan, et al. Quantum key distribution, Phys. Rev. A 79, 032341 (2016).
\bibitem{s9} R. Raussendorf, J. H. Briegel, A one-way quantum computer, Phys. Rev. Lett. 86, 5188 (2001).
\bibitem{s10} G. Gour, Family of concurrence monotones and its applications, Phys. Rev. A 71, 012318 (2005).
\bibitem{s11} W. K. Wootters, Entanglement of formation of an arbitrary state of two qubits, Phys. Rev. Lett. 80, 2245 (1998).
\bibitem{s12} C. Eltschka, J. Siewert, Quantifying entanglement resources, J. Phys. A: Math. Theor. 47, 424005 (2014).
\bibitem{s13} V. Coffman, J. Kundu, W. K. Wootters, Distributed entanglement, Phys. Rev. A 61, 052306 (2000).
\bibitem{s14} T. J. Osborne, F. Verstraete, General monogamy inequality for bipartite qubit entanglement, Phys. Rev. Lett. 96, 220503 (2006).
\bibitem{s15} Y. K. Bai, N. Zhang, M. Y. Ye, et al. Exploring multipartite quantum correlations with
the square of quantum discord, Phys. Rev. A 88, 012123 (2013).
\bibitem{s16} Y. K. Bai, Y.F. Xu, and Z.D. Wang, General Monogamy Relation for the Entanglement of Formation in Multiqubit Systems, Phys. Rev. Lett. 113, 100503 (2014).
\bibitem{s17} X. N. Zhu and S. M. Fei, Entanglement monogamy relations of qubit systems, Phys. Rev. A 90, 024304 (2014).
\bibitem{s18} Ou Y C, Fan H. Monogamy inequality in terms of negativity for three-qubit states,
Phys. Rev. A 75, 062308 (2007).
\bibitem{s19} J. S. Kim, A. Das, B. C. Sanders, Entanglement monogamy of multipartite higher-dimensional quantum systems using convex-roof extended negativity, Phys. Rev. A 79, 012329 (2009).
\bibitem{s20} H. He, G. Vidal, Disentangling Theorem and Monogamy for Entanglement Negativity, Phys. Rev. A 91, 012339 (2015)
\bibitem{s21} Y. Luo, Monogamy of $\alpha$th power entanglement measurement in qubit systems, Ann. Phys. 362, 511 (2015)

\bibitem{s22}
Z. X. Jin, S. M. Fei, Finer distribution of quantum correlations among multiqubit systems, Quantum Inf Process 18, 21
(2019).
\bibitem{s23} Z. X. Jin and S. M. Fei, Tighter entanglement monogamy relations of qubit systems, Quantum Inf. Process. 16. 77 (2017).
\bibitem{s24} Z. X. Jin, Jun. Li, S. M. Fei, et al. Tighter monogamy relations in multiqubit systems, Phys. Rev A 97, 032336 (2018).
\bibitem{s25} L. M. Yang, B. Chen, S. M. Fei, Z. X. Wang, Tighter constraints of multiqubit entanglement, Commun. Theor. Phys. 71 (2019).
\bibitem{s26} W. W. Liu, Z. F. Yang, S. M. Fei, Tighter monogamy and polygamy relations of quantum entanglement in multi-qubit systems, International Journal of Theoretical Physics 60, 4177 (2021).
\bibitem{s27} G. Gour, S. Bandyopadhyay, B. C. Sanders, Dual monogamy inequality for entanglement, Journal of Mathematical Physics 48, 012108 (2007).
\bibitem{s28} G. Gour, D. A. Meyer, B. C. Sanders, Deterministic entanglement of assistance and monogamy constraints, Phys. Rev. A 72, 042329 (2005).
\bibitem{s29} J. S. Kim, General polygamy inequality of multi-party quantum entanglement, Phys. Rev. A 85, 062302 (2012).
\bibitem{s30} J. S. Kim, Negativity and tight constraints of multiqubit entanglement. Phys. Rev. A 97, 012334 (2018).

\bibitem{s32} G. Vidal, R. F. Werner, Computable measure of entanglement, Phys. Rev. A 65, 032314 (2002).
\bibitem{s33} X. N. Zhu and S. M. Fei, Monogamy properties of qubit systems, Quantum Inf Process 18, 23 (2019).
\bibitem{s34} Z. X. Jin, S. M. Fei, C. F. Qiao, Polygamy relations of multipartite systems, Quantum Inf Process 18, 105 (2019).
\bibitem{s35} Z. X. Jin, S. M. Fei, C. F. Qiao, Complementary quantum correlations among multipartite systems, Quantum Inf Process 19, 101 (2020).
\bibitem{s36}  A. Uhlmann, Fidelity and concurrence of conjugated states. Phys. Rev. A 62, 032307 (2000).
\bibitem{s37} W. K. Wootters, Entanglement of formation of an arbitrary state of two qubits, Phys. Rev. Lett. 80, 2245 (1998).
\bibitem{s38} A. Acin, A. Andrianov, L. Costa, E. Jane, J. I. Latorre, R. Tarrach, Generalized Schmidt decomposition and classification of three-quantum-bit states, Phys. Rev. Lett. 85, 1560 (2000).
\bibitem{s39} Z. X. Jin, S. M. Fei, Superactivation of monogamy relations for nonadditive quantum correlation measures, Phys. Rev. A 99, 032343 (20s19).


\end{thebibliography}
\end{document}